\documentclass[conference]{IEEEtran}
\IEEEoverridecommandlockouts
\usepackage{amsmath,amssymb,amsfonts,makecell}
\usepackage{booktabs,multirow}
\usepackage{algorithmic}
\usepackage{graphicx}
\usepackage{textcomp}
\usepackage{xcolor}
\usepackage{stfloats}
\usepackage{hyperref}
\usepackage[table]{xcolor}
\usepackage{arydshln}
\usepackage[square,sort&compress,comma,numbers]{natbib}
\usepackage{algorithm}
\usepackage[caption=false]{subfig}
\usepackage{enumitem}
\newcommand{\comm}[1]{$\triangleright$ {\small \textit{#1}}}
\usepackage{tikz}
\usetikzlibrary{calc}

\definecolor{gray_utku}{rgb}{0.3, 0.3, 0.3}
\newcommand{\var}[1]{\textcolor{gray_utku}{\text{\tiny$\pm$#1}}}
\def\x{{\mathbf x}}
\def\y{{\mathbf y}}
\def\E{{\mathbf E}_\Omega}
\def\Et{{\mathbf E}_\Theta}
\def\El{{\mathbf E}_\Lambda}

\def\bepsilon{{\boldsymbol \epsilon}}

\makeatletter
\usepackage{xspace}
\def\@onedot{\ifx\@let@token.\else.\null\fi\xspace}
\DeclareRobustCommand\onedot{\futurelet\@let@token\@onedot}

\def\ie{\emph{i.e}\onedot}

\begin{document}

\title{Automated Tuning for Diffusion Inverse Problem Solvers without Generative Prior Retraining}

\author{\IEEEauthorblockN{Ya\c{s}ar Utku Al\c{c}alar$^{*\dagger}$ \qquad Junno Yun$^{*\dagger}$ \qquad Mehmet Ak\c{c}akaya$^{*\dagger}$}
\IEEEauthorblockA{$^*$Department of Electrical \& Computer Engineering, University of Minnesota, MN, USA}
\IEEEauthorblockA{$^\dagger$Center for Magnetic Resonance Research, University of Minnesota, MN, USA}}

\maketitle

\begin{abstract}
Diffusion/score-based models have recently emerged as powerful generative priors for solving inverse problems, including accelerated MRI reconstruction. While their flexibility allows decoupling the measurement model from the learned prior, their performance heavily depends on carefully tuned data fidelity weights, especially under fast sampling schedules with few denoising steps. Existing approaches often rely on heuristics or fixed weights, which fail to generalize across varying measurement conditions and irregular timestep schedules. In this work, we propose Zero-shot Adaptive Diffusion Sampling (ZADS), a test-time optimization method that adaptively tunes fidelity weights across arbitrary noise schedules without requiring retraining of the diffusion prior. ZADS treats the denoising process as a fixed unrolled sampler and optimizes fidelity weights in a self-supervised manner using only undersampled measurements. Experiments on the fastMRI knee dataset demonstrate that ZADS consistently outperforms both traditional compressed sensing and recent diffusion-based methods, showcasing its ability to deliver high-fidelity reconstructions across varying noise schedules and acquisition settings.

\end{abstract}

\begin{IEEEkeywords}
Artificial intelligence, diffusion models, zero-shot learning, computational imaging, MRI.
\end{IEEEkeywords}

\vspace{-1ex}
\section{Introduction}
\begin{figure}[t]
  \centering
  \subfloat[{\scriptsize Fast sampling strategy employing uniformly spaced steps.} \label{fig:uniform_sch}]{\includegraphics[width=0.91\columnwidth]{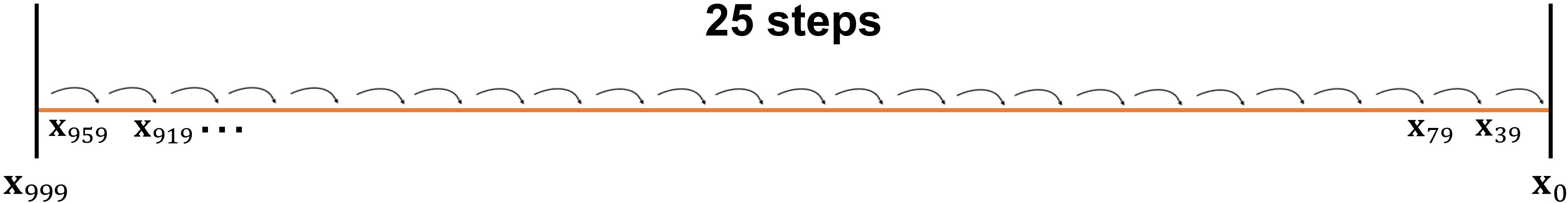}}\\
  \subfloat[{\scriptsize Fast sampling strategy using non-uniform (irregular) step sizes.} \label{fig:irregular_sch}]{\includegraphics[width=0.91\columnwidth]{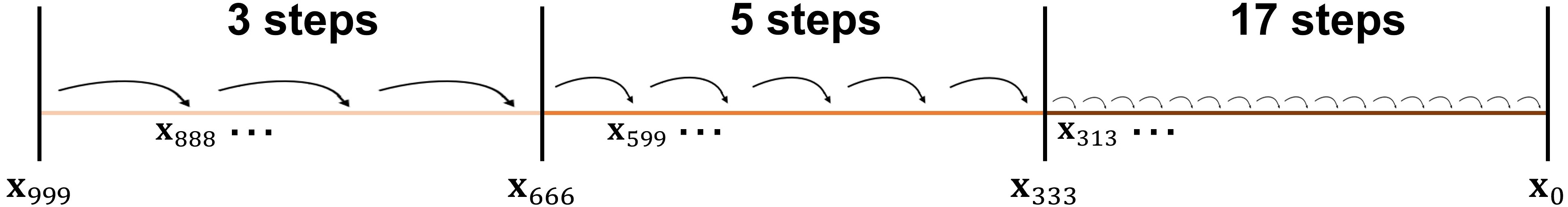}}
  \caption{Comparison of fast sampling strategies for diffusion models.}
  \label{fig:schedules}
  \vspace{-2ex}
\end{figure}

Generative models have seen rapid progress in recent years, with diffusion/score-based models emerging as a leading class, enabling high-quality data generation across modalities such as images, audio, and video~\cite{song2019generative,ho2020ddpm,kong2021diffwave,song2021ddim,ho2022video_DMs}. Beyond generation, diffusion models have also been applied as powerful priors for solving ill-posed inverse problems, achieving state-of-the-art results in both natural and medical imaging~\cite{song2022solving-medical,chung2023dps,wang2023ddnm,song2023pgdm}. Early techniques for accelerating the sampling process typically employed uniformly spaced denoising steps (Fig.\ref{fig:uniform_sch}), while later works proposed more advanced irregular sampling schedules that allocate more computation to low-noise regions where high-frequency details are recovered~\cite{dhariwal2021beatGANs,alcalar2024ZAPS} (Fig.~\ref{fig:irregular_sch}).

In medical imaging, particularly MRI, inverse problems arise from the need to accelerate acquisition by sampling only a subset of k-space. Physics-driven deep learning (PD-DL) methods have addressed this using unrolled optimization networks to map undersampled to fully sampled data~\cite{hammernik2018VarNet,aggarwal2019MoDL,hosseini2020dense}. However, these models often fail to generalize across acquisition settings due to their reliance on fixed forward models shaped by vendor, hardware, and protocol differences~\cite{yaman2022zeroshot}.

Diffusion-based reconstruction offers a compelling alternative by decoupling the prior from the measurement process~\cite{jalal2021robust,song2022solving-medical,chung2024decomposed}. This allows the same pretrained prior to adapt flexibly at inference to various forward operators, improving robustness across sampling patterns and scanner configurations. Despite this flexibility, their performance is still sensitive to data fidelity weights, which are often hand-tuned heuristically for varying noise levels and measurement signal-to-noise ratio (SNR). This sensitivity is especially problematic in regimes with very low numbers of function evaluations (NFEs), where irregular schedules are required to preserve fine details, but the corresponding fidelity weights are particularly difficult to hand-tune due to their non-uniform behavior across timesteps. While such schedules have been explored in the context of unconditional image generation~\cite{dhariwal2021beatGANs} and natural image restoration~\cite{alcalar2024ZAPS}, their role in accelerated MRI reconstruction remains largely unstudied.

\begin{tikzpicture}[remember picture,overlay]
  \node[anchor=north, align=justify, text width=0.83\paperwidth] 
  at ($(current page.south)+(0,21mm)$) 
  {\footnotesize \textcopyright~2025 IEEE. Personal use of this material is permitted. 
   Permission from IEEE must be obtained for all other uses, in any current or future media, 
   including reprinting/republishing this material for advertising or promotional purposes, 
   creating new collective works, for resale or redistribution to servers or lists, 
   or reuse of any copyrighted component of this work in other works.};
\end{tikzpicture}

\looseness=-1
To tackle these challenges, we introduce \textbf{Z}ero-shot \textbf{A}dap\-tive \textbf{D}iffusion \textbf{S}ampling (ZADS), a method that learns to adapt data fidelity weights across arbitrary noise schedules during inference. Instead of relying on fixed heuristics or retraining of the generative diffusion prior, ZADS treats the denoising process as a fixed unrolled sampler and optimizes the fidelity weights in a test-time self-supervised fashion. Experiments on multi-coil fastMRI knee data  show that ZADS improves reconstruction quality over conventional compressed sensing and fixed-weight diffusion methods such as diffusion posterior sampling (DPS) and decomposed diffusion sampling (DDS), even with the same or fewer NFEs. By jointly adapting to both the noise schedule and the measurement conditions, ZADS offers a flexible robust solution for diffusion-based computational MRI.

\section{Methods}
\subsection{MRI Inverse Problem and PD-DL Unrolling}
\begin{figure*}[!t]
    \centerline{\includegraphics[width=0.88\textwidth]{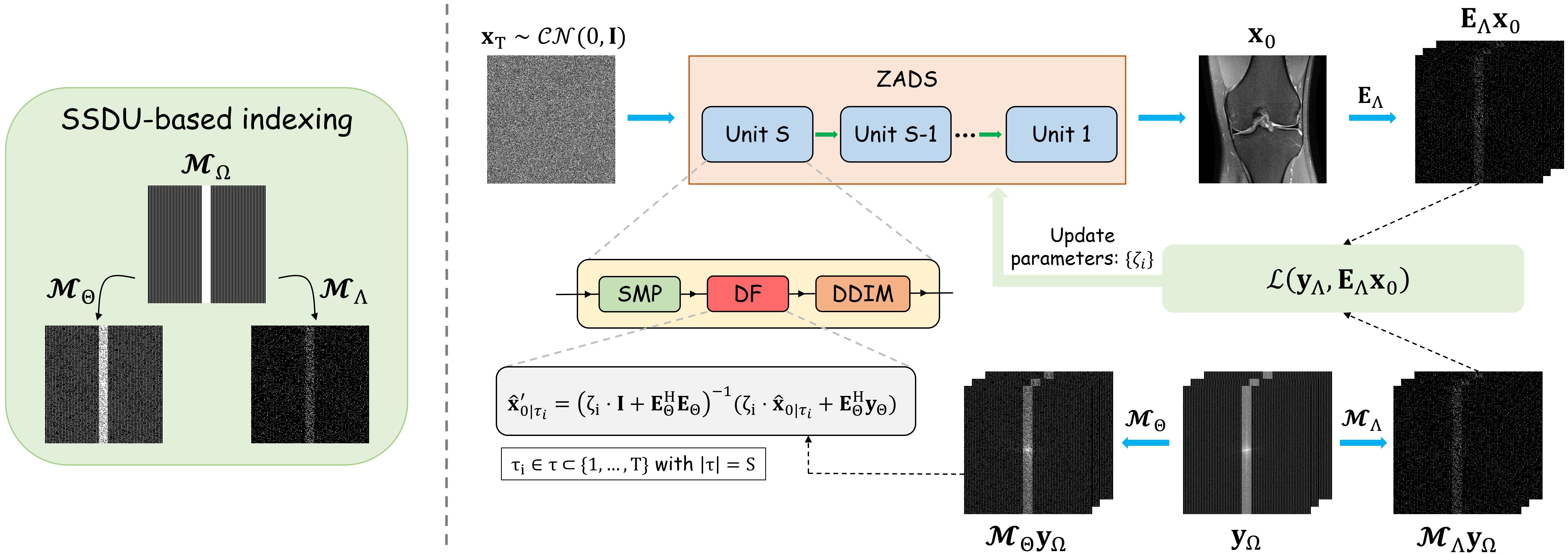}}
    \vspace{-.2cm}
    \caption{Our Zero-shot Adaptive Diffusion Sampling (ZADS) framework treats the diffusion sampling process as an unrolled architecture, where timestep-dependent data fidelity (DF) weights $\{\zeta_i\}$ are optimized at test time using SSDU loss. Here, the acquired k-space locations $\Omega$ are split into two disjoint sets: $\Theta$ for data fidelity updates, and $\Lambda$ held out for fidelity-weight tuning. SMP denotes the score model prediction, \ie, the Tweedie denoised estimate, $\hat{\bf x}_{0|\tau_i}$.}
    \label{fig:methods}
    \vspace{-4ex}
\end{figure*}

The inverse problem in MRI reconstruction solves:
\begin{equation} \label{eq:invMRI}
    \arg\min_{\x} \|\y_\Omega - \E\x\|_2^2 + \mathcal{R}(\x),
\end{equation}
where \(\Omega\) is the k-space sampling pattern, and \(\E\) is the associated multi-coil encoding operator incorporating Fourier undersampling and coil sensitivities. The first quadratic term in \eqref{eq:invMRI} enforces fidelity with measured data, and \(\mathcal{R}(\cdot)\) serves as a regularizer encoding prior information about the image. A commonly used PD-DL approach tackles this problem by unrolling traditional iterative algorithms \cite{fessler2020SPM, yaman2021_3D-LGE, akcakaya2022_SPMsurvey, alcalar2025_CUPID} into a fixed number of learnable stages and train the network end-to-end to simultaneously learn the proximal operator defined by \(\mathcal{R}(\cdot)\) and tune data fidelity weights. In supervised setups, the training objective minimizes the difference between the network output and the fully-sampled reference image~\cite{hammernik2018VarNet,aggarwal2019MoDL,knoll2020deep-survey,gu2022revisiting,heckel2024deep}. To overcome the difficulty of acquiring fully-sampled data in practical MRI settings, self-supervision via data undersampling (SSDU)~\cite{yaman2020SSDU,yaman2022mmssdu} proposes to divide the acquired k-space into two disjoint sets: $\Theta$ for data consistency and network input, and $\Lambda$ for supervision. The model is then trained to minimize the discrepancy between predicted and actual measurements over $\Lambda$, enabling learning directly from undersampled data:
\begin{equation}
    \min_{\boldsymbol{\theta}} \mathbb{E} \left[ \mathcal{L} \left( \mathbf{y}_{\Lambda}, \mathbf{E}_{\Lambda}(f(\mathbf{y}_{\Theta}, \mathbf{E}_{\Theta}; \boldsymbol{\theta})) \right) \right].
\end{equation}

\subsection{Diffusion Models}
Diffusion models are a class of generative models that synthesize data by reversing a gradual noising process. In the standard formulation, known as Denoising Diffusion Probabilistic Models (DDPM)~\cite{ho2020ddpm}, a clean image \(\x_0\) is progressively corrupted through a forward process over \(\mathrm{T}\) timesteps, producing noisy samples \(\x_t\). The forward process is defined as:
\begin{equation}
    q(\x_t | \x_0) = \mathcal{N}(\x_t; \sqrt{\bar{\alpha}_t} \x_0, (1 - \bar{\alpha}_t) \mathbf{I}),
\end{equation}
where \(\bar{\alpha}_t = \prod_{s=1}^t \alpha_s\), and \(\alpha_t = 1 - \beta_t\) with \(\beta_t\) representing a predefined noise schedule. A neural network \(\hat{\bepsilon}_t=\bepsilon_{\theta^*}(\x_t, t)\) is trained to approximate the noise \(\boldsymbol{\epsilon}\) added during the forward process, enabling the reverse process to iteratively denoise the sample. To accelerate sampling, Denoising Diffusion Implicit Models (DDIM)~\cite{song2021ddim} introduce a deterministic and non-Markovian alternative to DDPM. At each timestep \(t\), DDIM estimates the original clean image using Tweedie's formula~\cite{efron2011tweedie}:
\begin{equation}
    \hat{\mathbf{x}}_{0|t} = (\mathbf{x}_t - \sqrt{1 - \bar{\alpha}_t} \hat{\bepsilon}_t)/ \sqrt{\bar{\alpha}_t}. \label{eq:tweedie}
\end{equation}
The next sample \(\x_{t-1}\) is then computed as:
\begin{align}
    \x_{t-1} &= \sqrt{\bar{\alpha}_{t-1}} \hat{\x}_{0|t} + \sqrt{1 -\bar{\alpha}_{t-1} - \sigma_t(\eta)^2} \hat{\bepsilon}_t +\sigma_t(\eta)\mathbf{z}, \nonumber \\
    &\triangleq \sqrt{\bar{\alpha}_{t-1}} \hat{\x}_{0|t} + \tilde{\boldsymbol{\omega}}_t,
\end{align}
where \(\boldsymbol{z} \sim \mathcal{N}(0, \mathbf{I})\) and \(\eta \in [0,1]\) controls the level of stochasticity. Setting \(\eta = 0\) yields a purely deterministic process, whereas \(\eta = 1\) corresponds to DDPM sampling.

\subsection{Diffusion Generative Model Inverse Problem Solvers}
Diffusion-based inverse problem solvers aim to generate reconstructions that both match measured data and lie on the learned data manifold. A prominent example, DPS~\cite{chung2023dps}, alternates DDIM-based denoising with a gradient correction step that enforces data consistency while constraining updates to remain on the same noisy manifold. Specifically, at each step $t$, it updates the sample via:
\begin{equation}
    \x_{t-1} = \sqrt{\bar{\alpha}_{t-1}}\hat{\mathbf{x}}_{0|t} - \zeta\cdot \nabla_{\x_t}\ell(\hat{\mathbf{x}}_{0|t}) + \tilde{\boldsymbol{\omega}}_t,
\end{equation}
where $\hat{\x}_{0|t}$ is the denoised estimate obtained via~\eqref{eq:tweedie}, $\ell(\cdot)$ denotes the data consistency loss (\ie, $\| \y - \E\hat{\mathbf{x}}_{0|t} \|_2^2$), and $\zeta$ is a heuristically tuned step size parameter. Building on DPS, DDS~\cite{chung2024decomposed} proposes a geometry-aware refinement strategy that avoids direct gradient updates by leveraging linear solvers. The key insight in DDS is to assume that the clean data manifold \( \mathcal{M} \) is locally affine or well-approximated by its tangent space \( \mathcal{T}_t \) at the denoised estimate \( \hat{\mathbf{x}}_{0|t} \). Under this assumption, the Jacobian \( \partial \hat{\mathbf{x}}_{0|t} / \partial \mathbf{x}_t \) reduces to \( \mathcal{P}_{\mathcal{M}} / \sqrt{\bar{\alpha}_t} \), yielding:
\begin{equation}
    \hat{\x}_{0|t} - \zeta\cdot \nabla_{\x_t}\ell(\hat{\x}_{0|t})= \mathcal{P}_{\mathcal{M}}\left(\hat{\x}_{0|t} - \zeta^\prime \nabla_{\hat{\x}_{0|t}} \ell(\hat{\x}_{0|t})\right),
\end{equation}
where \( \mathcal{P}_{\mathcal{M}} \) denotes the projection operator onto the tangent space. This insight motivates replacing the single projected step with classical conjugate gradient (CG) steps constrained to the Krylov subspace. Since the Krylov subspace spans the tangent space of $\mathcal{M}$ at $\hat{\x}_{0|t}$, DDS performs an M-step CG update within this subspace to obtain the refined estimate:
\begin{equation}
    \hat{\mathbf{x}}^\prime_{0|t} = \mathrm{CG}(\E^\mathrm{H}\E +\zeta\cdot\mathbf{I}, \E^\mathrm{H}\y+\zeta\cdot\hat{\mathbf{x}}_{0|t}, \hat{\mathbf{x}}_{0|t}, \mathrm{M}),
\end{equation}
where $\zeta$ is the data fidelity weight selected through heuristic tuning. The final sample at timestep \( t-1 \) is computed as:
\begin{equation}
    \mathbf{x}_{t-1} = \sqrt{\bar{\alpha}_{t-1}} \hat{\mathbf{x}}^\prime_{0|t} + \tilde{\boldsymbol{\omega}}_t.
\end{equation}

\subsection{Proposed Zero-shot Adaptive Diffusion Sampling (ZADS)}
While recent diffusion-based solvers have improved data-consistent diffusion sampling, they still depend on fixed noise schedules and manually selected fidelity weights~\cite{alcalar2024ZAPS}. This reliance may result in suboptimal reconstructions, since the optimal weighting often depends on the measurement SNR or noise level, which varies across MRI acquisition settings. Furthermore, this limits the use of irregular noise schedules, as heuristically choosing the fidelity weights becomes impractical due to each timestep affecting the output differently.
\begin{algorithm}[!t]
  \caption{Zero-shot Adaptive Diffusion Sampling (ZADS)} \label{alg:algo}
  \begin{algorithmic}[1]
    \REQUIRE $\bepsilon_{\theta^*},\; \mathrm{T},\; \{\alpha_t\}_{t=1}^\mathrm{T},\; \eta,\; \mathbf{E}_\Omega,\; \mathbf{y}_\Omega,\;\mathrm{M}$
    %\vspace{.5ex}
    \STATE $\mathbf{x}_\mathrm{T} \sim \mathcal{CN}(\mathbf{0}, \mathbf{I})$ 
    \STATE \comm{Selection of an irregular schedule}
    \STATE $\tau \subset \{1,...,\mathrm{T}\}$ extending over a length of $\mathrm{S}<\mathrm{T}$
    %\vspace{.5ex}
    %\STATE \comm{SSDU-based index splitting}
    \STATE Get $\{\mathbf{E}_{\{\Theta,\;\Lambda\}},\;\mathbf{y}_{\{\Theta,\;\Lambda\}}\}$ \hfill \comm{SSDU-based index splitting}
    %\vspace{.5ex}
    \FOR {epoch $\textbf{in}\text{ epochs}$}
        \FOR {$i = \mathrm{S},...,1$}
            \STATE $\hat{\bepsilon}_{\tau_i} \gets \bepsilon_{\theta^*}(\mathbf{x}_{\tau_i}, \tau_i)$
            \vspace{.25ex}
            \STATE \comm{Score model prediction (Tweedie denoising)}
            \vspace{.25ex}
            \STATE $\hat{\mathbf{x}}_{0|\tau_i} \gets (\mathbf{x}_{\tau_i} - \sqrt{1 - \bar{\alpha}_{\tau_i}} \hat{\bepsilon}_{\tau_i}) / \sqrt{\bar{\alpha}_{\tau_i}}$
            \vspace{.5ex}
            \STATE \comm{Data consistency}
            \vspace{.5ex}
            \STATE $\mathbf{E}_{\text{CG}} \gets \zeta_i \cdot \mathbf{I} + \mathbf{E}_\Theta^\mathrm{H} \mathbf{E}_\Theta$
            %\vspace{.5ex}
            \STATE $\mathbf{y}_{\text{CG}} \gets \zeta_i \cdot \hat{\mathbf{x}}_{0|\tau_i} + \mathbf{E}_\Theta^\mathrm{H} \mathbf{y}_\Theta$
            %\vspace{.5ex}
            \STATE $\hat{\mathbf{x}}_{0|\tau_i}^{\prime} \gets \text{CG}(\mathbf{E}_{\text{CG}}, \mathbf{y}_{\text{CG}}, \hat{\mathbf{x}}_{0|\tau_i}, \mathrm{M})$
            \vspace{.7ex}
            \STATE \comm{DDIM sampling}
            \vspace{.7ex}
            \STATE $\mathbf{z} \sim \mathcal{CN}\mathbf{(0, I)} \text{ if } \tau_i>1$, else $\mathbf{z=0}$
            \vspace{.5ex}
            \STATE $\sigma_{\tau_i} \gets \eta \sqrt{\frac{(1 - \bar{\alpha}_{\tau_{i-1}})}{(1 - \bar{\alpha}_{\tau_i})} \left(1 - \frac{\bar{\alpha}_{\tau_i}}{\bar{\alpha}_{\tau_{i-1}}}\right)}$
            %\vspace{.5ex}
            \STATE {\small $\x_{{\tau_{i-1}}} \gets \sqrt{\bar{\alpha}_{\tau_{i-1}}} \hat{\x}_{0|\tau_i}^{\prime} + \sqrt{1 -\bar{\alpha}_{\tau_{i-1}} - \sigma_{\tau_i}^2} \boldsymbol{\hat{\epsilon}}_{\tau_i} +\sigma_{\tau_i}\mathbf{z}$}
            \vspace{.3ex}
            %\STATE $\mathbf{z} \sim \mathcal{CN}(\mathbf{0}, \mathbf{I})$
            %\STATE $\mathbf{x}_{t-1} \gets \sqrt{\bar{\alpha}_{\tau_{i-1}}} \hat{\mathbf{x}}_0^{'} - \sqrt{1 - \bar{\alpha}_{\tau_{i-1}} - \eta^2 \tilde{\beta}_{\tau_i}^2} \hat{\bepsilon}_{\tau_i} + \eta \tilde{\beta}_{\tau_i}\mathbf{z}$
        \ENDFOR
        %\vspace{.3ex}
        %\STATE \comm{No stochasticity or data-consistency at the last step}
        %\vspace{.5ex}
        %\STATE $\mathbf{x}_0 \gets (\mathbf{x}_1 - \sqrt{1 - \bar{\alpha}_1} \bepsilon_{\theta^*}(\mathbf{x}_1)) / \sqrt{\bar{\alpha}_1}$
        %\STATE \comm{Update network parameters $\{\zeta_i\}$}
        %\STATE $\mathcal{L} \gets \mathrm{MSE}(\y_\Lambda,\bf{A}_\Lambda\x_0)$
        \STATE Update network parameters $\{\zeta_i\}$ via $\mathcal{L}(\y_\Lambda,\mathbf{E}_\Lambda\x_0)$
        %\vspace{.3ex}
    \ENDFOR
    \RETURN $\mathbf{x}_0$
  \end{algorithmic}
\end{algorithm}

To address these limitations, we propose ZADS, a unified framework that adaptively sets timestep-dependent data fidelity weights for each timestep to better align the posterior with the observed measurements, without retraining the unconditional diffusion model. Inspired by algorithm unrolling in PD-DL~\cite{monga2021algorithm,demirel2021EMBC_20fold_7TfMRI,hammernik2023SPM,alcalar2024_ISBI,alcalar2025_SPIC_SSDU}, ZADS treats the diffusion sampling procedure in DDS as a fixed unrolled process and optimizes only the fidelity weights across timesteps, without modifying the underlying score model. Unlike previous approaches that rely on fixed heuristics, ZADS learns these weights at test time by minimizing a self-supervised loss derived from held-out measurements. Specifically, we adopt a strategy inspired by SSDU, wherein the acquired k-space is split into disjoint sets: one subset is used to enforce data consistency during CG, while the other is held out for fidelity-weight optimization.

For an arbitrary noise schedule $\tau \subset \{1, \ldots, \mathrm{T}\}$, with $S = |\tau| \ll \mathrm{T}$, ZADS obtains the refined denoised estimate through:
\begin{equation}
    \hat{\mathbf{x}}^\prime_{0|\tau_i} = \mathrm{CG}(\Et^\mathrm{H}\Et +\zeta_i\cdot\mathbf{I}, \Et^\mathrm{H}\y+\zeta_i\cdot\hat{\mathbf{x}}_{0|\tau_i}, \hat{\mathbf{x}}_{0|\tau_i}, \mathrm{M}).
\end{equation}
After producing the final output $\x_0$ using only a few NFEs, a physics-driven loss is computed on the held-out $\Lambda$ samples:
\begin{equation}
    \mathcal{L}(\y_\Lambda,\mathbf{E}_\Lambda\x_0) = \frac{||\y_\Lambda-\El\x_0||_1}{||\y_\Lambda||_1}+\frac{||\y_\Lambda-\El\x_0||_2}{||\y_\Lambda||_2}. \label{eq:loss}
\end{equation}
Fig.~\ref{fig:methods} provides a high-level illustration of our algorithm.

We note the distinction between proposed ZADS and our earlier zero-shot approximate posterior sampling (ZAPS) method~\cite{alcalar2024ZAPS}, which optimizes log-likelihood weights using all acquired measurements for natural image restoration. While such a strategy is effective for DPS-type sampling, as it applies mild corrections that remain close to the noisy manifold, the stronger CG-based updates in DDS may lead to overfitting if the same measurements are used for both inference and supervision, necessitating the use of hold-out masking in the current setup. The complete sampling algorithm is outlined in Algorithm~\ref{alg:algo}.

\section{Experimental Evaluation}
\subsection{Imaging Experiments and Implementation Details}
\begin{table}[!t]
    \setlength{\tabcolsep}{2pt}
    \caption{Quantitative comparison of reconstruction methods under R=4 equispaced undersampling for Coronal PD and Coronal PD-FS knee MRI. Best: \textbf{BOLD}, second-best: \underline{Underlined}}
    \begin{center}
    \begin{tabular}{@{}p{2cm}cccccc@{}}
    \hline
    \addlinespace[3pt]
    \multirow{2}{*}{Method} 
    & \multicolumn{2}{c}{Cor PD, Knee MRI} 
    & \multicolumn{2}{c}{Cor PD-FS, Knee MRI} \\
    %& \multirow{2}{*}{Inference Time (s)} \\
    \cmidrule(lr){2-3} \cmidrule(lr){4-5}
    & PSNR [dB] $\uparrow$ & SSIM $\uparrow$ & PSNR [dB] $\uparrow$ & SSIM $\uparrow$ & \\
    \hline
    \addlinespace[3pt]
    %\makecell[l]{Compressed \\ Sensing~\cite{lustig2007sparse}} & xx.xx\var{x.xx} & 0.xxx\var{x.0xx} & xx.xx\var{x.xx} & 0.xxx\var{x.0xx} \\%& x.xx\\
    $\ell_1$-Wavelet~\cite{lustig2007sparse} & 31.35\var{2.83} & 0.881\var{0.033} & 28.58\var{2.51} & 0.683\var{0.095} \\%& x.xx\\
    \addlinespace[3pt]
    DPS (1000)~\cite{chung2023dps} & 34.90\var{2.65} & 0.891\var{0.034} & 30.68\var{3.86} & 0.743\var{0.096} \\%& x.xx\\
    \addlinespace[3pt]
    DDS (25)~\cite{chung2024decomposed} & 32.60\var{2.18} & 0.896\var{0.023} & \underline{30.76\var{3.19}} & \underline{0.793\var{0.068}} \\%& x.xx\\
    \addlinespace[3pt]
    DDS (250)~\cite{chung2024decomposed} & \underline{34.93\var{2.20}} & \underline{0.899\var{0.038}} & 28.45\var{3.87} & 0.658\var{0.125} \\%& x.xx\\
    \addlinespace[3pt]
    \arrayrulecolor{gray} \hdashline
    \addlinespace[3pt]    
    ZADS {\bf (Ours)} & \textbf{36.32\var{2.08}} & \textbf{0.938\var{0.021}} & \textbf{32.48\var{2.95}} & \textbf{0.818\var{0.063}} \\%& \textbf{x.xx} \\
    \addlinespace[3pt]
    \arrayrulecolor{black} \hline
    \end{tabular}
    \vspace{-4ex}
    \label{tab:table}
    \end{center}
\end{table}
We used the NYU fastMRI multi-coil knee dataset~\cite{knoll2020fastmri_dataset-journal}, which contained coronal proton density (cor PD) and coronal proton density with fat suppression (cor PD-FS) scans, acquired at a matrix size of 320$\times$320 using 15 coils. We applied retrospective uniform undersampling to both datasets using an acceleration factor of $R=4$, retaining 24 central k-space lines. Our experiments focused on equidistant sampling schemes, which are standard in clinical MRI and produce structured aliasing artifacts that are considerably harder to suppress than the noise-like artifacts introduced by random sampling~\cite{knoll2019assessment}.

\begin{figure*}[!t]
    \centerline{\includegraphics[width=0.92\textwidth]{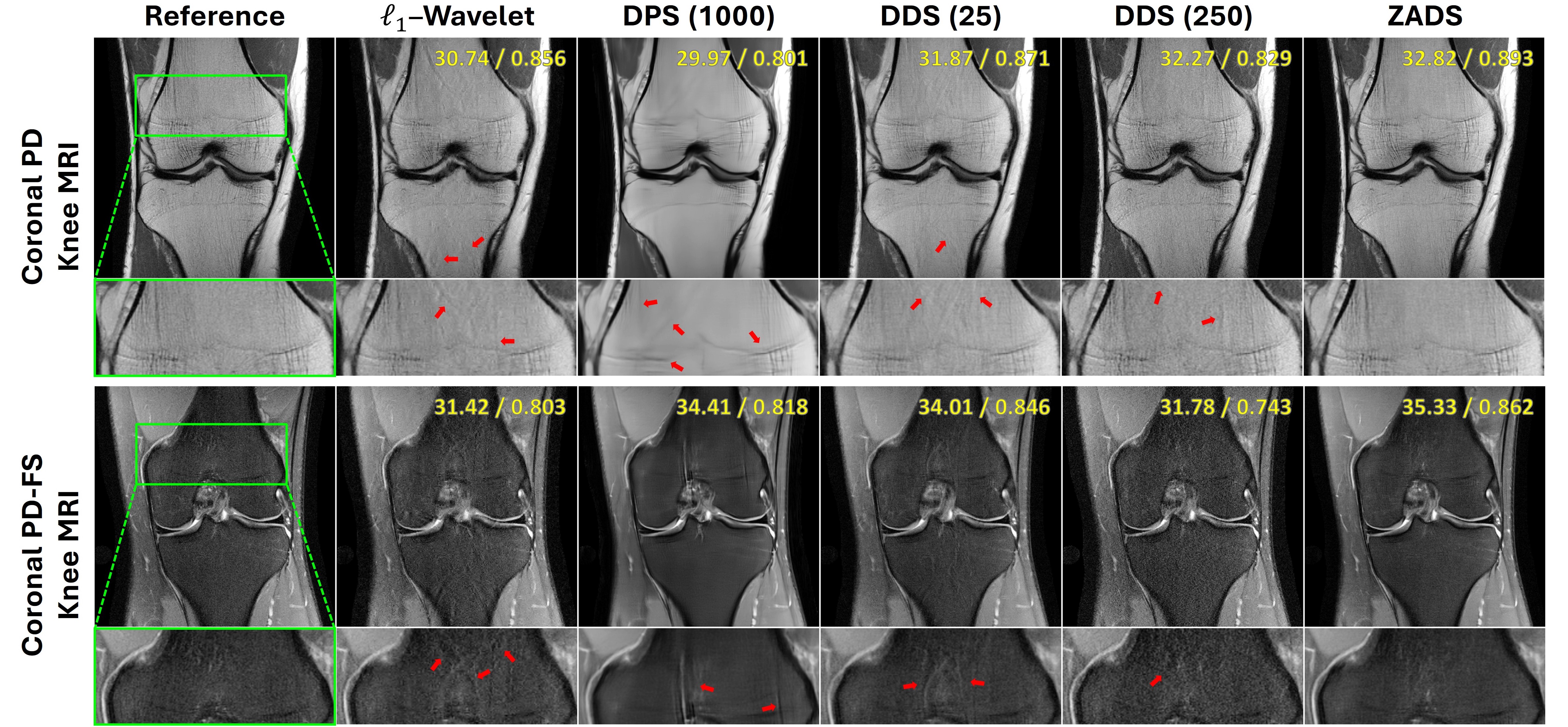}}
    \caption{Representative reconstructions from the coronal PD and PD-FS datasets ($R=4$, equidistant). DPS exhibits blurring and artifacts, while DDS shows either residual artifacts (25 steps) or noise amplification (250 steps). ZADS produces the most faithful reconstructions, effectively reducing noise and artifacts.}
    \label{fig:results}
    \vspace{-2ex}
\end{figure*}

\begin{figure}[!b]
    \vspace{-2ex}
    \centerline{\includegraphics[width=0.99\columnwidth]{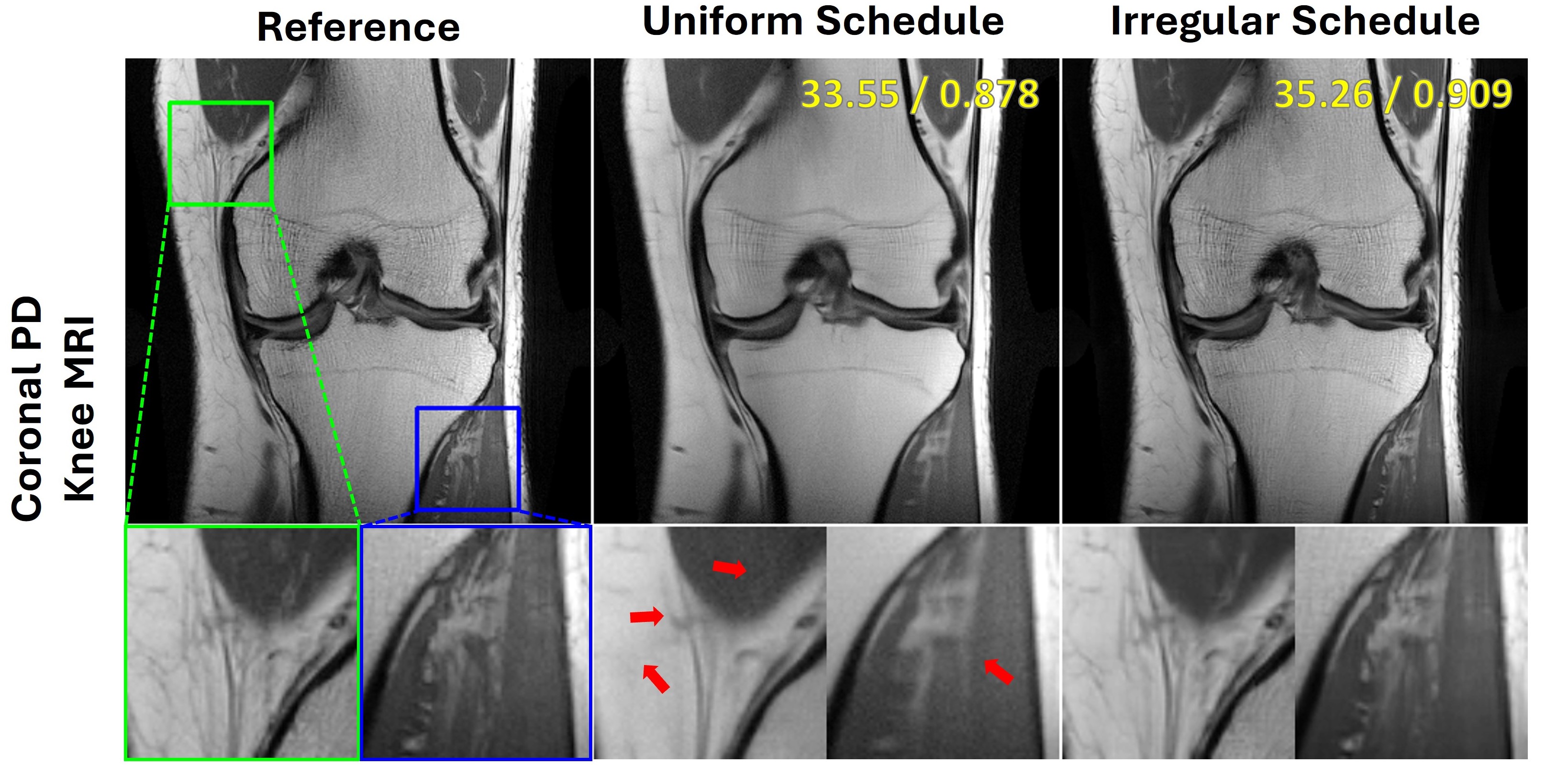}}
    \caption{Comparison of uniform and irregular noise schedules within the ZADS framework. Irregular schedules more effectively preserve fine structural details, resulting in sharper reconstructions.}
    \label{fig:ablation}
\end{figure}

We used a pretrained unconditional diffusion model from~\cite{chung2024decomposed}, trained on FastMRI knee images, without any additional retraining. For evaluation, we selected 100 central slices from 10 subjects per dataset, resulting in a total of 200 slices. Undersampled k-space data were generated by applying the sampling mask $\Omega$ to the \emph{noisy} fully-sampled measurements. For quantitative evaluation, we used 25 sampling steps following a ``17,5,3'' schedule given in Fig.~\ref{fig:irregular_sch}, with 10 fine-tuning epochs, resulting in a total of 250 NFEs. Following \cite{yaman2020SSDU}, we set the sampling ratio $\rho = |\Lambda| / |\Omega|$ to 0.4.

We compared our method against conventional $\ell_1$-wavelet compressed sensing, as well as two diffusion-based baselines: DPS and DDS. Both diffusion methods were implemented using their official public repositories. To ensure a fair comparison, all methods used the same pretrained unconditional diffusion model and employed DDIM sampling with $\eta = 0.85$. We used 15 CG iterations for both DDS and ZADS.

\subsection{Results}
Fig.~\ref{fig:results} presents reconstruction comparisons on the coronal PD and PD-FS datasets among $\ell_1$-wavelet compressed sensing, DPS with 1000 sampling steps, DDS with 25 and 250 steps, and ZADS using 25 steps fine-tuned over 10 epochs. The $\ell_1$-wavelet method produces visibly inferior reconstructions, failing to recover fine structures in both contrasts. While DPS benefits from a large number of sampling steps, it produces overly smoothed outputs and exhibits notable residual artifacts, particularly in the PD-FS case. The DDS (25) results further highlight the importance of combining irregular sampling schedules with adaptive fidelity weight tuning. Despite using the same number of sampling steps as ZADS, DDS exhibits visible artifacts, whereas ZADS produces cleaner reconstructions, demonstrating the benefits of test-time adaptation. On the other hand, the importance of adapting fidelity weights to the underlying SNR becomes evident from the DDS (250) results. This configuration performs reasonable on the high-SNR coronal PD dataset, however, it also amplifies noise on the low-SNR PD-FS dataset, indicating that a fixed regularization weight does not generalize effectively across different noise levels. ZADS achieves the most effective artifact and noise suppression across both datasets, as further supported by the quantitative results in Table~\ref{tab:table}.

Finally, our ablation study in Fig.~\ref{fig:ablation} demonstrates that within the ZADS framework, irregular sampling schedules capture fine structural details more effectively than uniform schedules, leading to visibly sharper reconstructions.

\begingroup
\renewcommand\thefootnote{}
\footnotetext{This work was partially supported by NIH R01HL153146, NIH R01EB032830, NIH P41EB027061.}
\addtocounter{footnote}{-1}
\endgroup

\section{Discussion and Conclusion}
In this study, we introduced ZADS, a novel framework that adaptively tunes data fidelity weights across arbitrary diffusion noise schedules at test time without retraining the generative prior. Similar to our early work that explored algorithm unrolling for diffusion model-based inverse problem solvers for the first time~\cite{alcalar2024ZAPS}, ZADS treats the sampling process as a fixed unrolled architecture and optimizes timestep-dependent weights using a self-supervised loss on held-out k-space measurements. By avoiding retraining the diffusion model, unlike recent works~\cite{kamilov2025diff_unfolding}, we leverage the strengths of the diffusion prior across different SNRs, while providing automatic adaptability for data fidelity. Experiments on the fastMRI knee dataset demonstrate that ZADS consistently outperforms other diffusion-based methods, particularly when using irregular timesteps that better preserve fine details. Future work will explore automated noise schedule design and a more systematic analysis of the SSDU split ratio $\rho$.

\fontsize{9}{11}\selectfont
\bibliographystyle{IEEEbib}
\bibliography{refs}

\end{document}